\documentclass[conference]{IEEEtran}
\usepackage{algorithm}        
\IEEEoverridecommandlockouts
\usepackage{cite}
\usepackage{amsmath,amssymb,amsfonts}
\usepackage{algorithmic}
\usepackage{graphicx}
\usepackage{textcomp}
\usepackage{xcolor}
\usepackage{subfig}
\usepackage[hidelinks]{hyperref}
\def\BibTeX{{\rm B\kern-.05em{\sc i\kern-.025em b}\kern-.08em
    T\kern-.1667em\lower.7ex\hbox{E}\kern-.125emX}}
\begin{document}

\title{Traffic-Aware Domain Partitioning and Load-Balanced 
Inter-Domain Routing for LEO Satellite Networks\\
}

\author{\IEEEauthorblockN{Chen Zhou}
\IEEEauthorblockA{\textit{School of Communications and} \\
\textit{Information Engineering} \\
\textit{Chongqing University of Posts and} \\
\textit{Telecommunications}\\
Chongqing, China \\
s240132075@stu.cqupt.edu.cn}
\and
\IEEEauthorblockN{Jiangtao Luo}
\IEEEauthorblockA{\textit{School of Communications and} \\
\textit{Information Engineering} \\
\textit{Chongqing University of Posts and} \\
\textit{Telecommunications}\\
Chongqing, China \\
luojt@cqupt.edu.cn\thanks{ Jiangtao Luo and Yongyi Ran are the
corresponding authors. (Email: luojt@cqupt.edu.cn,
ranyy@cqupt.edu.cn). This work is supported by the National
Natural Science Foundation of China (No.~U25B2033, No.~U23A20275
 and No.~62171072).}}
\and
\IEEEauthorblockN{Yongyi Ran}
\IEEEauthorblockA{\textit{School of Communications and} \\
\textit{Information Engineering} \\
\textit{Chongqing University of Posts and} \\
\textit{Telecommunications}\\
Chongqing, China \\
ranyy@cqupt.edu.cn}
}

\maketitle

\begin{abstract}
Low Earth Orbit (LEO) satellite networks provide global coverage
and low latency, yet high node mobility, uneven traffic distribution,
and stochastic link failures pose severe challenges for inter-domain
routing. Existing approaches either neglect graph-structured topology
or lack dynamic awareness of real-time link states, struggling to
balance load distribution and routing reliability. This paper proposes
DTAR, a traffic-aware deep reinforcement learning approach for
inter-domain routing in LEO satellite networks. A multi-objective
NSGA-II algorithm first generates an offline domain partition
maximizing intra-domain traffic ratio and minimizing load imbalance.
A Graph Attention Network dynamically encodes inter-domain
link traffic intensity, load distribution, and fault status, upon
which an action-masked PPO agent learns routing decisions online. Simulations on a 288-satellite Walker constellation
against multiple baselines demonstrate that DTAR
significantly reduces link load imbalance and end-to-end
delay, while improving routing success rate and reducing
packet loss rate across normal, traffic surge, and fault
scenarios.
\end{abstract}


\section{Introduction}
\IEEEPARstart{W}{ith} the rapid deployment of large-scale
Low Earth Orbit (LEO) satellite constellations such as
Starlink and OneWeb, LEO satellite networks have emerged
as a critical component of future 6G communications due
to their low propagation delay and wide coverage \cite{b1}.
To efficiently manage inter-satellite link resources in
large-scale constellations, satellite networks are
typically partitioned into domains, with inter-domain
routing protocols coordinating cross-domain traffic
scheduling. As the core function of the satellite network
control plane, inter-domain routing directly determines
end-to-end latency, load balancing capability, and fault
resilience.

However, inter-domain routing in LEO networks faces three
fundamental challenges: frequent topology changes, spatiotemporally
uneven traffic distribution, and stochastic link failures.
Existing methods fall short in addressing these challenges.
Traditional shortest-path algorithms such as Dijkstra are
load-oblivious and prone to hotspot congestion.
Q-learning-based methods such as QRLSN~\cite{b2} suffer from
poor Q-table generalization in large-scale multi-domain topologies.
ELB~\cite{b3} adopts load-weighted shortest-path routing for
non-geostationary satellite networks, yet relies on static
link weights without real-time traffic state awareness.
Domain partitioning-based methods have been actively
explored: SHORT~\cite{b4} organizes LEO networks into
stable hierarchical routing domains based on
orbital-geodetic coordinates, and Eunomia~\cite{b5}
leverages spectral clustering for mobility-aware domain
assignment. CDPAR~\cite{b6} combines domain partitioning
with DQN routing for adaptive routing decisions.
However, these methods do not sufficiently account for
traffic distribution during partitioning, and lack
real-time inter-domain link state awareness.

Graph neural networks and deep reinforcement learning
have shown strong potential for LEO routing optimization:
Rao et al. combined deep graph attention with
incremental evolutionary strategies for dynamic LEO
routing \cite{b7}, Zhang et al. proposed GRLR jointly
optimizing routing for mega-constellations with GNN and
reinforcement learning \cite{b8}, Li et al. proposed
a multi-agent hierarchical DRL approach for multi-domain
collaborative satellite routing \cite{b9}, Xu et al. integrated GNN and DRL for inter-satellite
routing with load balancing objectives \cite{b10}, and
Xia et al. proposed distributed DRL for joint computing
and routing in LEO constellations \cite{b11}. Nevertheless,
none of these works explicitly model domain-based structures
with real-time inter-domain link state encoding for
load-balanced routing.

To address these limitations, this paper proposes DTAR
(Deep Traffic-Aware inter-domain Routing),
a traffic-aware deep reinforcement learning approach for
load-balanced inter-domain routing in LEO satellite
networks. A two-stage framework decouples offline
traffic-aware domain partitioning from online adaptive
routing decision-making. Extensive simulations on a
288-satellite Walker constellation demonstrate that DTAR
outperforms four baseline methods across normal, traffic
surge, and link failure scenarios.
The main contributions are as follows:
\vspace{-0.3em}
\begin{enumerate}
\setlength{\itemsep}{0pt}
\setlength{\parskip}{0pt}
\setlength{\parsep}{0pt}
\item A multi-objective NSGA-II-based domain partitioning
method simultaneously maximizes intra-domain traffic ratio
and minimizes inter-domain load imbalance, providing a
structured topological foundation for routing.
\item A GAT-based domain-level state encoder explicitly
encodes inter-domain link traffic intensity, load
distribution, and fault status for real-time perception
of dynamic network conditions.
\item An action-masked PPO-based routing policy  ensures routing feasibility under topology
dynamics, link congestion, and fault scenarios.
\end{enumerate}

\section{System Model}

\subsection{Network Model}

A LEO satellite constellation forms a network modeled as an
undirected graph $\mathcal{G}_s=(\mathcal{V}_s, \mathcal{E}_s)$,
where $\mathcal{V}_s=\{v_1, v_2, \ldots, v_N\}$ denotes the set
of satellite nodes and $\mathcal{E}_s$ the set of inter-satellite
links (ISLs). The constellation is distributed across $P$ orbital
planes with $S$ satellites per plane. Under the standard 4-ISL
topology, each satellite $v_i$ maintains links to its two
intra-plane neighbors and two inter-plane counterparts.
An edge $e_{i,j}=e_{j,i}\in\mathcal{E}_s$ represents the
physical ISL between $v_i$ and $v_j$, characterized at time
step $t$ by link load $w_{ij}(t)$ and fault status
$b_{ij}(t)\in\{0,1\}$.

To reduce routing complexity in large-scale constellations,
satellites are grouped into $K$ domains, forming a domain-level
graph $\mathcal{G}=(\mathcal{V},\mathcal{E})$, where
$\mathcal{V}=\{D_1,\ldots,D_K\}$ is the set of domain nodes
and $\mathcal{E}$ is the set of inter-domain edges.
An edge $(D_i,D_j)\in\mathcal{E}$ exists if at least one
physical ISL connects satellites in the respective domains.
Each domain $D_k$ comprises a set of geographically adjacent
satellites and is characterized at time step $t$ by:
normalized intra-domain satellite count $n_k$,
traffic load $L_k(t)$, cross-domain link count $c_k$,
intra-domain fault node ratio $f_k(t)$, and surge hotspot
indicator $s_k(t)\in\{0,1\}$.
Note that $c_k$ is time-invariant, as the domain partition
is fixed offline and the inter-domain edge set $\mathcal{E}$
does not change during online routing.
Each inter-domain edge $(D_i,D_j)\in\mathcal{E}$ carries 
three dynamic attributes: available link ratio $a_{ij}(t)$, 
inter-domain link load $W_{ij}(t)$, and fault status flag 
$B_{ij}(t)\in\{0,1\}$.
When $B_{ij}(t)=1$, all cross-domain ISLs between $D_i$ and
$D_j$ are simultaneously failed, i.e., $a_{ij}(t)=0$, and
routing between these two domains is blocked. When
$B_{ij}(t)=0$, the inter-domain link is fully operational.
The domain partition is generated offline by NSGA-II and
remains static throughout online routing.

\begin{figure}[t]
\centering
\includegraphics[width=0.75\columnwidth]{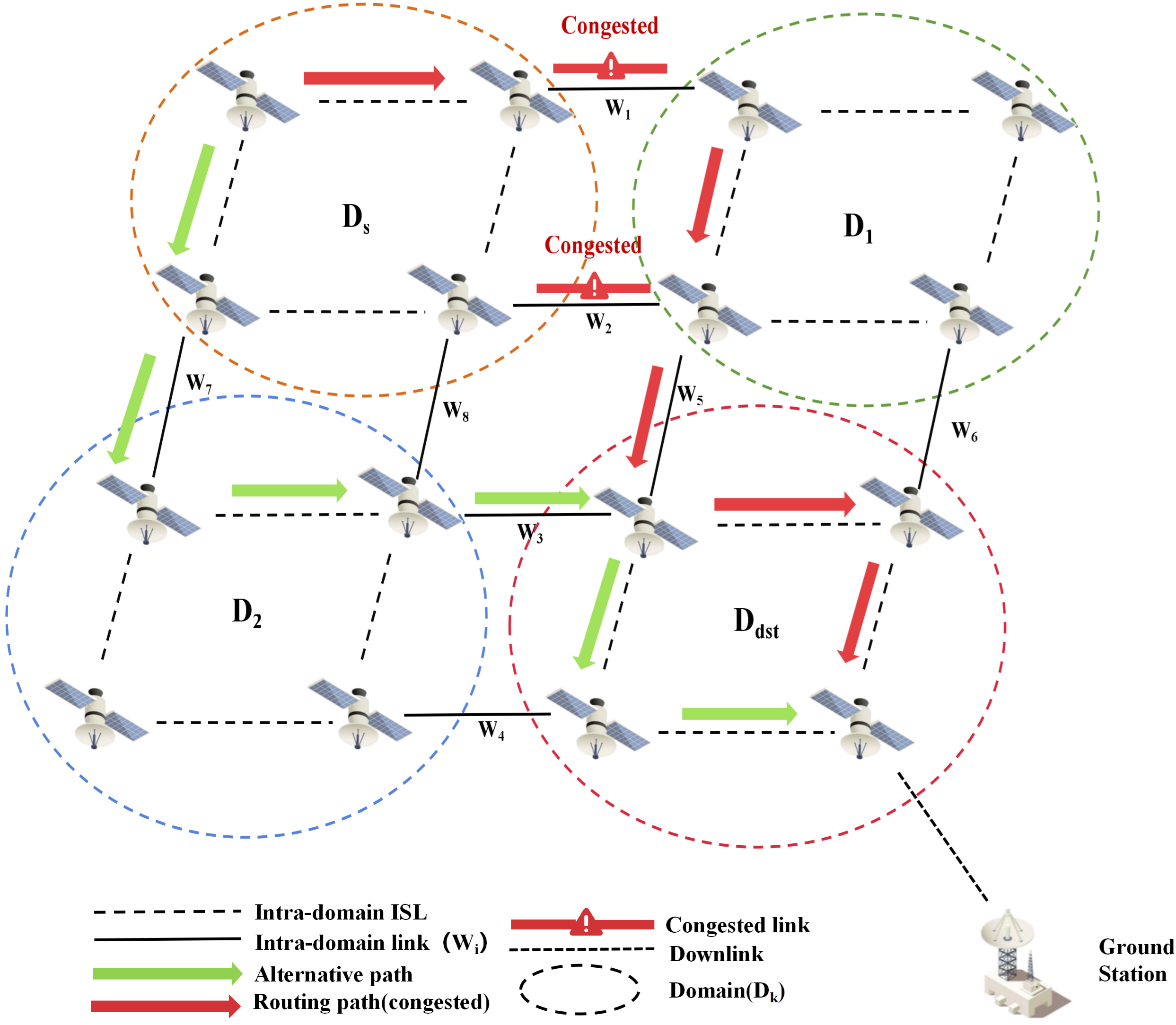}
\caption{The LEO satellite inter-domain routing problem: 
load-balanced path selection across partitioned domains.}
\label{fig:system_model}
\end{figure}

\subsection{Traffic and Link State Model}

Traffic demands within the constellation follow a
spatiotemporally uneven distribution reflecting the peak-hour
patterns of terrestrial hotspot regions. At each time step,
a daily-pattern generator produces the satellite-level traffic
matrix $\mathbf{T}(t)\in\mathbb{R}^{N\times N}$, where
$T_{ij}(t)$ denotes the traffic intensity from satellite $v_i$
to $v_j$ at time $t$. The aggregated traffic load of domain
$D_k$ is defined as:
\begin{equation}
  L_k(t) = \sum_{i\in D_k}\!\left(
    \sum_{j=1}^{N} T_{ij}(t) + \sum_{j=1}^{N} T_{ji}(t)
  \right)
\end{equation}
which aggregates bidirectional traffic associated with all
satellites in domain $D_k$.
In the traffic surge scenario, a random cross-domain link pair
is selected as the hotspot and the traffic of satellites within
its surrounding region is amplified by a factor of $\mu$ to
emulate burst load conditions.

Link failures are injected at the inter-domain edge level as
a stochastic process: each inter-domain link fails independently
at each time step with probability $p_f$ and recovers with
probability $p_r$. Fault injection is constrained to preserve
the connectivity of $\mathcal{G}$, ensuring routing feasibility
under all fault scenarios.

\subsection{Problem Formulation}

For a routing request from source domain $D_s$ to destination
domain $D_{\mathrm{dst}}$, the objective is to find a
domain-level path
$\mathcal{P}=\{D_s, D_{h_1}, \ldots, D_{\mathrm{dst}}\}$
in $\mathcal{G}$ that minimizes inter-domain link load
imbalance and maximizes routing success rate $\mathrm{SR}(t)$,
where $\mathrm{SR}(t)$ denotes the ratio of successfully
delivered routing requests to the total number of requests
at time step $t$.
Load imbalance is measured by the coefficient of variation
(CV) of inter-domain link loads:
\begin{equation}
  \mathrm{CV}(t)=\frac{\mathrm{std}(W_{ij}(t))}
  {\mathrm{mean}(W_{ij}(t))},\quad(D_i,D_j)\in\mathcal{E}
\end{equation}
where $\mathrm{CV}(t)$ is defined when
$\mathrm{mean}(W_{ij}(t))>0$; in the degenerate case of
zero network load, $\mathrm{CV}(t)$ is set to~$0$.

The optimization problem is formulated as:
\begin{equation}
\min_{\mathcal{P}}\ \mathrm{CV}(t),\quad
\max_{\mathcal{P}}\ \mathrm{SR}(t)
\end{equation}
\[
\mathrm{s.t.}\quad|\mathcal{P}|-1\leq H_{\max},\quad
\forall\,(D_i,D_j)\in\mathcal{P}:\ B_{ij}(t)=0
\]
where $D_s$ and $D_{\mathrm{dst}}$ are connected in
$\mathcal{G}$ under the current fault state.

The two objectives are jointly addressed within the Markov 
Decision Process (MDP) framework: $\mathrm{CV}(t)$ and 
$\mathrm{SR}(t)$ are incorporated into a scalar reward function 
through a weighted combination of a load balancing term and 
terminal success/failure rewards, enabling the DRL agent to 
optimize both objectives simultaneously via policy gradient updates.

Due to the dynamic nature of the domain-level
topology---including traffic fluctuations, stochastic link
failures, and surge events---this problem is intractable for
conventional optimization methods in real time. This problem is therefore formulated as an MDP, and deep reinforcement learning is employed for online decision-making, 
as detailed in Section~\ref{sec:method}.

\section{Domain Partitioning and Inter-Domain Routing}
\label{sec:method}

This section presents the DTAR framework for load-balanced inter-domain routing in LEO satellite networks. DTAR adopts a two-stage design: in the offline stage, NSGA-II generates a traffic-aware domain partition that provides a structured topological foundation for routing; in the online stage, a GAT-PPO agent perceives real-time inter-domain link states and makes adaptive routing decisions. The two stages address optimization problems at different timescales, decoupling the combinatorial complexity of domain partitioning from the real-time dynamics of routing decisions.
Within each domain, intra-domain forwarding follows shortest-path routing based on hop count, which is orthogonal to the inter-domain routing problem addressed in this work.

\begin{figure*}[t]
\centering
\includegraphics[width=0.75\textwidth]{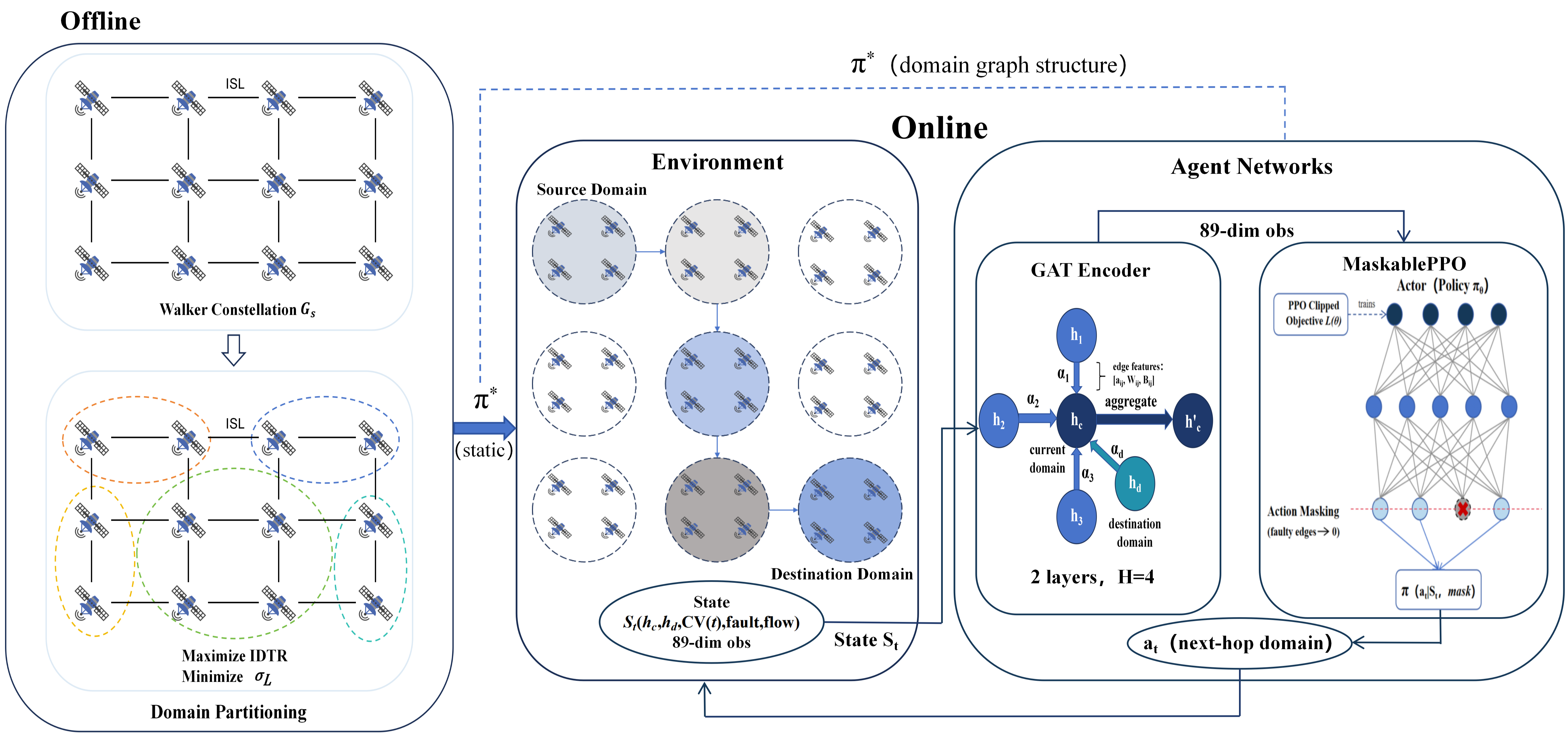}
\caption{The proposed DTAR framework.}
\label{fig:system}
\end{figure*}

\subsection{Traffic-Aware Domain Partitioning via NSGA-II}

Given the satellite topology $\mathcal{G}_s$ and traffic
matrix $\mathbf{T}$, the partitioning problem seeks an
assignment $\pi:\mathcal{V}_s\to\{0,\ldots,K-1\}$ that
jointly optimizes two objectives.

\textbf{Objective 1: Maximizing Intra-Domain Traffic Ratio
(IDTR).}
\begin{equation}
  \mathrm{IDTR}(\pi)=
  \frac{\sum_{i}\sum_{j}
        T_{ij}\cdot\mathbf{1}[\pi(i)=\pi(j)]}
       {\sum_{i}\sum_{j} T_{ij}}
\end{equation}
A higher IDTR confines more traffic within domains, reducing
cross-domain routing overhead.

\textbf{Objective 2: Minimizing Inter-Domain Load Deviation.}
The aggregated load of domain $\mathcal{D}_d$ is:
\begin{equation}
  L_d = \sum_{i\in\mathcal{D}_d}\!\left(
        \sum_{j} T_{ij} + \sum_{j} T_{ji}\right)
\end{equation}
The load deviation $\sigma_L=\mathrm{std}(L_0,\ldots,L_{K-1})$
is minimized to achieve balanced load distribution across
domains.

NSGA-II~\cite{b11} is applied to solve this bi-objective
problem, with each individual encoding $\pi$ as a vector
of $N$ domain labels. The algorithm maintains a population
of $M$ individuals and evolves them over $G$ generations
via non-dominated sorting and crowding-distance selection,
preserving solution diversity along the Pareto front.
Single-point crossover exchanges domain assignment segments
between parent individuals to explore new partition
configurations. Three mutation modes are applied
stochastically:
\textit{intra-orbit perturbation} (60\%) preserves traffic
locality by reassigning satellites to the domain of an
orbitally adjacent neighbor;
\textit{traffic-aware migration} (20\%) moves a border node
from the most-loaded domain to the least-loaded to reduce
$\sigma_L$;
\textit{random boundary exploration} (20\%) maintains
population diversity.

After each genetic operation, a multi-stage repair operator
enforces hard constraints through the following sequential
stages:
(1)~\textit{empty-domain revival}: splits nodes from the
largest domain to repopulate any empty domain;
(2)~\textit{upper-bound repair}: migrates boundary nodes
from oversized domains to adjacent ones;
(3)~\textit{lower-bound repair}: borrows boundary nodes
from neighboring large domains to undersized ones;
(4)~\textit{connectivity repair}: reassigns nodes in
disconnected components to the nearest adjacent domain;
(5)~\textit{post-connectivity lower-bound re-check}:
re-applies lower-bound repair with donor eligibility
relaxed to 2-hop neighbors to handle residual undersized
domains after connectivity repair.
The operator enforces domain size within $[n_{\min},
n_{\max}]$ and intra-domain connectivity, leaving IDTR
and $\sigma_L$ fully subject to NSGA-II optimization.
The final partition is selected from the Pareto front by
maximizing a normalized score with equal weight on IDTR
improvement and $\sigma_L$ reduction.

\subsection{GAT-based Domain State Encoding}

To enable real-time perception of dynamic inter-domain
topology states, we design a GAT-based domain-level state
encoder that jointly encodes inter-domain link traffic
intensity, load distribution, and fault status into
structured embeddings.

\textbf{Input Features.}
Each domain node $v\in\mathcal{V}$ carries a
\textbf{6-dimensional} feature vector comprising:
normalized intra-domain satellite count $n_k$,
domain traffic load $L_k(t)$,
normalized cross-domain link count $c_k$,
intra-domain fault node ratio $f_k(t)$,
normalized domain load index $\ell_k(t)$,
and surge hotspot indicator $s_k(t)$.
Each inter-domain edge $(D_i,D_j)\in\mathcal{E}$ carries a
3-dimensional feature vector comprising: available link
ratio $a_{ij}(t)$, inter-domain link load $W_{ij}(t)$,
and fault status flag $B_{ij}(t)$.

\textbf{Architecture.}
The encoder consists of an input projection layer followed
by two GAT layers~\cite{b13} with hidden dimension 64,
$H{=}4$ attention heads, and output dimension 32. In each
layer, attention coefficients are computed jointly from
node and edge features. Let the pre-activation score be:
\begin{equation}
\footnotesize
  \alpha_{ij} =
  \mathrm{softmax}_i\!\left(
    \mathrm{LeakyReLU}\!\left(
      \mathbf{w}_a^\top\!
      [\mathbf{W}_n\mathbf{h}_i\|
       \mathbf{W}_n\mathbf{h}_j\|
       \mathbf{W}_e\mathbf{f}_{ij}]
    \right)
  \right)
\end{equation}
where $\mathbf{f}_{ij}$ is the edge feature vector and $\|$
denotes concatenation. Node representations are updated via
weighted aggregation with a residual connection and
LayerNorm for training stability:
\begin{equation}
\footnotesize
  \mathbf{h}_v^{(l)} =
  \mathrm{LayerNorm}\!\left(
    \mathrm{ELU}\!\left(
      \sum_{u\in\mathcal{N}(v)}\!
      \alpha_{uv}\mathbf{W}_n^{(l)}\mathbf{h}_u^{(l-1)}
    \right)
    +\mathbf{h}_v^{(l-1)}
  \right)
\end{equation}

\textbf{State Representation.}
The 89-dimensional PPO observation is formed by
concatenating: current domain embedding
$\mathbf{h}_c\in\mathbb{R}^{32}$, destination domain
embedding $\mathbf{h}_d\in\mathbb{R}^{32}$, a 3-dimensional
flow state (normalized current domain index, destination
index, and elapsed hops), an 18-dimensional neighbor
distance vector encoding shortest-hop distances to the
destination, and a 4-dimensional global state (time
progress, link load CV, surge indicator, fault indicator).

\subsection{Action-Masked PPO for Online Routing}

In the online routing stage, DTAR employs
an action-masked PPO~\cite{b14} to solve hop-by-hop routing
decisions on the domain-level graph $\mathcal{G}$.
Compared with conventional DRL approaches, the action
masking mechanism explicitly guarantees routing feasibility
under topology dynamics and link failures, eliminating
gradient waste on infeasible actions.

\textbf{Reward Function.}
The per-step reward comprises a direction shaping term,
a per-hop penalty, and terminal rewards:
\begin{equation}
\footnotesize
\begin{aligned}
  R(c,a,c') =\;
  &\delta\cdot(d(c,D_{\mathrm{dst}})
              -d(c',D_{\mathrm{dst}}))
   -\beta \\
  &+\mathbf{1}[c'=D_{\mathrm{dst}}]\cdot(r_s-\beta\cdot h)
   -r_f\cdot\mathbf{1}[\text{failure}]
\end{aligned}
\end{equation}
where $d(\cdot,D_{\mathrm{dst}})$ denotes the shortest-hop
distance to the destination in $\mathcal{G}$, $\delta$ is
the direction shaping coefficient, $\beta$ is the per-hop
penalty, $h$ is the number of hops taken upon arrival,
$r_s$ is the arrival reward, and $r_f$ is the failure
penalty. The arrival reward $r_s - \beta\cdot h$
incentivizes not only successful delivery but also shorter
paths, jointly encouraging high success rate and low
average path hops.

Load balancing is promoted through two complementary
mechanisms: the NSGA-II partition structurally confines
traffic within domains, while the GAT encoder exposes
real-time link loads $W_{ij}(t)$ as state inputs for
implicit congestion-aware path selection.

\textbf{Action Masking.}
Prior to each decision step, two categories of actions are
masked: (1)~neighboring domains whose every cross-domain
ISL is simultaneously faulty, ensuring physical
reachability; (2)~neighboring domains whose shortest-hop
distance to the destination exceeds the remaining hop
budget $H_{\max}-h_t$, enforcing the hop constraint
$|\mathcal{P}|-1\leq H_{\max}$ at the execution level.

\textbf{Training Strategy.}
Each episode independently samples a scenario
(normal\,/\,surge\,/\,fault) without curriculum scheduling.
Hyperparameters are listed in Table~\ref{tab:hparam} and
the training procedure is summarized in
Algorithm~\ref{alg:ppo}.
{\small
\begin{algorithm}[t]
\caption{Training Procedure of Action-Masked PPO for Inter-Domain
         Routing}
\label{alg:ppo}
\begin{algorithmic}[1]
\REQUIRE Domain partition $\pi^*$, domain-level graph
         $\mathcal{G}$, total timesteps $T$, rollout buffer
         size $N_{\mathrm{steps}}$, mini-batch size $B$,
         update epochs $E$
\ENSURE  Trained routing policy $\theta^*$
\STATE Initialize policy $\pi_\theta$ and value network
       $V_\phi$ with random weights
\STATE Build domain graph $\mathcal{G}$ and initialize
       GAT encoder with edge index from $\pi^*$
\STATE Initialize rollout buffer $\mathcal{B}$
\FOR{each training iteration}
    \STATE Sample scenario $\omega \sim
           \{\text{normal},\,\text{surge},\,\text{fault}\}$
    \STATE Reset environment, obtain initial observation
           $o_0$
    \FOR{$t = 1$ \TO $N_{\mathrm{steps}}$}
        \STATE Encode domain graph via GAT to obtain
               embeddings $\{\mathbf{h}_v\}$
        \STATE Construct observation $o_t$ by concatenating
               $\mathbf{h}_c$, $\mathbf{h}_d$, flow state,
               neighbor distances, global state
        \STATE Compute valid action mask $\mathbf{m}_t$
               by checking fault status and remaining
               hop budget of each neighboring domain
        \STATE Sample action $a_t \sim
               \pi_\theta(\cdot \mid o_t, \mathbf{m}_t)$
        \STATE Execute $a_t$, observe $o_{t+1}$ and reward
               $R_t$
        \STATE Store $(o_t, a_t, R_t, o_{t+1},
               \mathbf{m}_t)$ in $\mathcal{B}$
    \ENDFOR
    \STATE Compute advantages $\hat{A}_t$ via GAE
    \FOR{$e = 1$ \TO $E$}
        \FOR{each mini-batch of size $B$ from $\mathcal{B}$}
            \STATE Compute clipped PPO objective
                   $\mathcal{L}(\theta)$ and update
                   $\theta \leftarrow \theta -
                   \nabla_\theta\mathcal{L}(\theta)$
            \STATE Update $\phi$ via value loss
                   $\mathcal{L}_V(\phi)$
        \ENDFOR
    \ENDFOR
    \STATE Clear rollout buffer $\mathcal{B}$
\ENDFOR
\RETURN $\theta^* = \theta$
\end{algorithmic}
\end{algorithm}
}
\section{Simulation Results}

\subsection{Simulation Setup}

A system-level simulator is developed in Python using
NetworkX for topology modeling, PyTorch for the DRL network,
and Stable-Baselines3 for training, supporting
mixed-scenario sampling across normal, surge, and fault
conditions. Detailed simulation parameters are listed in
Table~\ref{tab:hparam}.
{\renewcommand{\thefootnote}{\relax}%
\footnote{Source code is available at 
\url{https://github.com/ChenZ-code/DTAR_Routing}.}}

\begin{table}[t]
\caption{Simulation Parameters}
\label{tab:hparam}
\centering
\begin{tabular}{|l|r|}
\hline
\textbf{Parameter} & \textbf{Value} \\
\hline
No. of satellites $N$            & 288               \\
No. of orbital planes $P$        & 12                \\
Satellites per plane $S$         & 24                \\
Orbital altitude                 & 1{,}450~km        \\
Inclination                      & $89^\circ$        \\
No. of domains $K$               & 18                \\
Population size $M$              & 100               \\
Generations $G$                  & 200               \\
Steps per episode                & 144               \\
Flows per step                   & 3                 \\
Max hops $H_{\max}$              & 9                 \\
Fault probability $p_f$          & 0.02              \\
Recovery probability $p_r$       & 0.02              \\
Surge multiplier $\mu$           & $5\times$         \\
Learning rate                    & $3\times10^{-4}$  \\
Total timesteps                  & $2.5\times10^{6}$ \\
Mini-batch size                  & 256               \\
Entropy coefficient              & 0.03              \\
\hline
\end{tabular}
\end{table}

The following four baseline methods are adopted for
comparison. \textbf{Dijkstra} is a classic shortest-path
algorithm using hop count as cost, oblivious to link load
states and prone to hotspot congestion under heavy traffic.
\textbf{ELB}~\cite{b3} is an explicit load-balancing routing
protocol for non-geostationary satellite networks that uses
load-weighted shortest-path routing, but relies on static
link weights without real-time traffic state awareness.
\textbf{QRLSN}~\cite{b2} is a Q-learning-based routing
method for LEO networks whose Q-table scale limits
generalization in large-scale multi-domain topologies.
\textbf{CDPAR}~\cite{b6} is a DQN-based adaptive routing
method combined with domain partitioning, but lacking
real-time inter-domain link state encoding.

All DRL baselines are trained for $2.5\times10^6$ steps
under the same mixed-scenario schedule and evaluated over
100 episodes. All quantitative results reported in the
following subsections are averaged over 100 evaluation
episodes. Evaluation metrics include the coefficient
of variation of inter-domain link loads (CV, lower is
better), end-to-end delay approximated as the number of
inter-domain hops multiplied by average per-hop
propagation delay~(ms), packet loss rate~(PLR, defined
as the ratio of failed routing requests to total requests),
and routing success rate~(SR, defined as the ratio of
successfully delivered routing requests to total requests).

\subsection{Comparison with Baselines}

\begin{figure}[t]
\centering
\subfloat[]{\includegraphics[width=0.48\columnwidth]{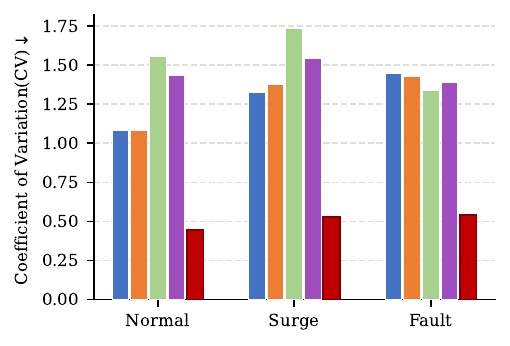}%
\label{fig:cv}}
\hfil
\subfloat[]{\includegraphics[width=0.48\columnwidth]{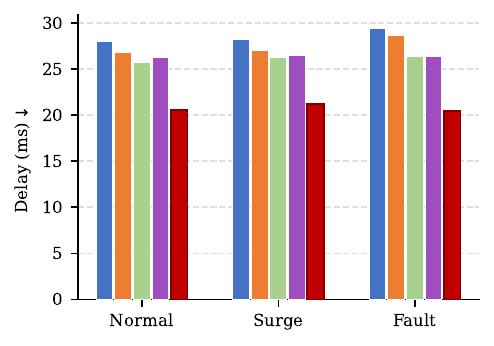}%
\label{fig:delay}}
\\
\subfloat[]{\includegraphics[width=0.48\columnwidth]{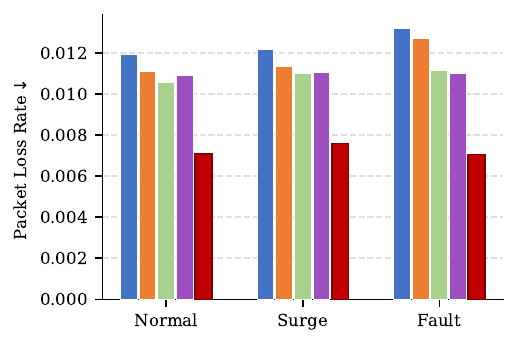}%
\label{fig:plr}}
\hfil
\subfloat[]{\includegraphics[width=0.48\columnwidth]{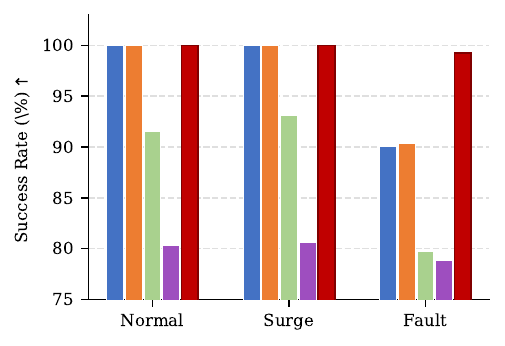}%
\label{fig:sr}}
\\
\includegraphics[width=0.90\columnwidth]{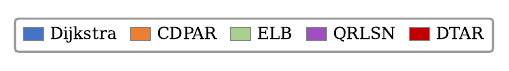}
\caption{Performance comparison of routing algorithms under
normal, surge, and fault scenarios.}
\label{fig:main}
\end{figure}

Fig.~\ref{fig:main} presents CV, end-to-end delay,
packet loss rate, and routing success rate of all methods
under normal, surge, and fault scenarios.

\textbf{Normal scenario.}
DTAR significantly reduces CV compared to all baselines.
The GAT encoder continuously encodes real-time inter-domain
link loads, guiding the agent to steer flows toward less
congested paths. Notably, ELB and QRLSN produce CV values
even higher than Dijkstra, demonstrating that
load-spreading heuristics without domain-level topology
awareness are counterproductive in multi-domain routing.
DTAR also achieves the lowest end-to-end delay owing to
its congestion-aware path selection.

\textbf{Traffic surge scenario.}
Under a $5\times$ localized traffic surge, DTAR achieves
the lowest CV across all methods. The surge hotspot
indicator in the node feature vector enables the agent
to detect congested domains and reroute flows before
congestion propagates. CDPAR, despite adopting domain
partitioning, performs comparably to Dijkstra due to its
lack of real-time link state encoding. The reduction in
CV is accompanied by a corresponding decrease in
end-to-end delay under surge conditions.

\textbf{Link failure scenario.}
DTAR improves routing success rate by 9.25 and 8.89
percentage points over Dijkstra and CDPAR, respectively.
The action masking mechanism eliminates routing attempts
over fully failed inter-domain edges, while the fault
status edge feature guides the agent toward resilient
alternative paths. ELB and QRLSN exhibit success rates
below 80\% due to the absence of explicit fault-aware
path selection, resulting in significantly higher packet
loss rates. Across all scenarios, DTAR consistently
reduces end-to-end delay and packet loss rate compared
to all baselines.

\subsection{Ablation Study}
\begin{figure}[t]
\centering
\includegraphics[width=0.75\columnwidth]{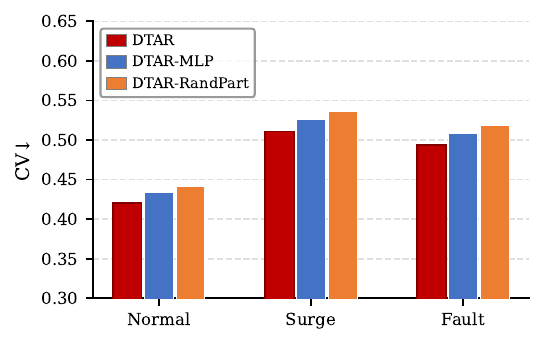}
\caption{Ablation study: CV comparison across three
scenarios.}
\label{fig:ablation}
\end{figure}

To isolate the contribution of each component, two ablation
variants are evaluated: \textbf{DTAR-MLP} replaces the GAT
encoder with a multilayer perceptron, and
\textbf{DTAR-RandPart} substitutes the NSGA-II partition
with a uniform orbital partition. Both variants are trained
under identical conditions to DTAR.
Fig.~\ref{fig:ablation} reports CV across all three scenarios.
Removing either component consistently degrades load balancing
performance across all scenarios. NSGA-II-based partitioning
plays a dominant role under fault scenarios, as traffic-aware
domain boundaries structurally limit cross-domain congestion
under topology disruption. GAT-based encoding contributes
more under surge conditions, where adaptive perception of
real-time link states enables timely rerouting around
congested domains. These results confirm that the two
components are complementary and both essential to DTAR's
robustness.

\subsection{Training Convergence}
\begin{figure}[t]
\centering
\includegraphics[width=0.75\columnwidth]{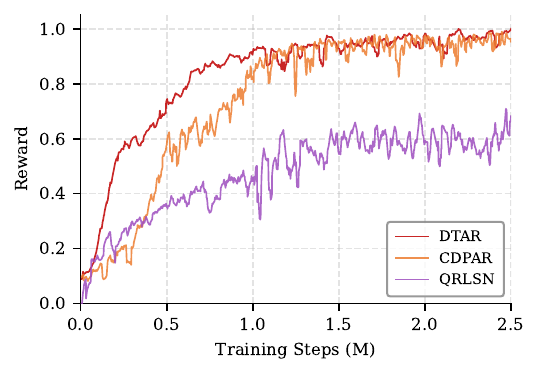}
\caption{Training convergence comparison.}
\label{fig:convergence}
\end{figure}

Fig.~\ref{fig:convergence} shows training reward curves of
DTAR, CDPAR, and QRLSN. DTAR converges stably within
approximately $1.5\times10^6$ steps with the highest final
reward. CDPAR converges more slowly without graph-structured
encoding, while QRLSN exhibits the slowest convergence,
limited by Q-table generalization in large-scale topologies.

\section{Conclusion}
This paper proposes DTAR, a traffic-aware deep reinforcement
learning framework for load-balanced inter-domain routing in
LEO satellite networks. NSGA-II generates an offline
traffic-aware domain partition, while a GAT encoder and
action-masked PPO agent realize online adaptive routing. 
Simulations on a 288-satellite Walker constellation
demonstrate significant improvements in load balance, routing
success rate, and end-to-end delay over four baselines across
normal, surge, and fault scenarios. Two limitations remain:
DTAR operates solely at the domain level without jointly
optimizing intra-domain paths, and generalization to
heterogeneous constellations remains to be verified.
Future work will address both directions.

\end{document}